\documentclass{article} \usepackage{amsmath} \usepackage{amssymb}
\usepackage{epsfig}
\newcommand{\cC}{\mathcal{C}}
\newcommand{\cA}{\mathcal{A}}

\newcommand{\cM}{\mathcal{M}}

\newcommand{\eg}{\textit{e.g. }}
\newcommand{\ie}{\textit{i.e. }}
\newcommand{\etc}{\textit{etc.}}

\def\SetOf#1#2{\left\{ #1  \,|\, #2 \right\} }

\font\openface=msbm10 at10pt

\def\Integers      {{\hbox{\openface Z}}}
\def\Reals         {{\hbox{\openface R}}}

\begin{document}
\title {The causal set approach to quantum gravity}

\author{Joe
Henson\footnote{Institute for Theoretical
Physics, University of Utrecht, Minnaert Building, Leuvenlaan 4, 3584 CE
Utrecht, The Netherlands.}}
\maketitle

\begin{abstract}
The ideas of spacetime discreteness and causality are important in
several of the popular approaches to quantum gravity. But if
discreteness is accepted as an initial assumption, conflict with
Lorentz invariance can be a consequence.  The causal set is a
discrete structure which avoids this problem and provides a
possible history space on which to build a ``path integral'' type
quantum gravity theory.  Motivation, results and open problems are
discussed and some comparisons to other approaches are made.  Some
recent progress on recovering locality in causal sets is
recounted.
\end{abstract}

\vskip 1cm

\textit{On the occasion of his 60th birthday, this article is
dedicated to Rafael Sorkin, without whom the causal set idea would
not have survived its infancy.}

 \vskip 1cm

How can we reach a theory of quantum gravity?  Many answers to
this question have been proposed, motivated by different
experiences and views of current theories \cite{Oriti:2006}. A
more specific set of questions might be: what demands should we
put on our framework, so that it is best able to meet all the
challenges involved in creating a theory of quantum gravity?  What
choices are most likely to give the correct theory, according to
the clues we have from known physics? Are there any problems with
our initial assumptions that may lead to trouble further down the
road?  The latter seems to be one of the most important strategic
questions when beginning to formulate a candidate theory.  For
example, can a canonical approach overcome the multifaceted
problem of time? And how far can a theory based on a fixed
background spacetime be pushed?  On the one hand, these questions
may only be answered in the very attempt to formulate the theory.
On the other, many such attempts have been made, and now that
quantum gravity research has built up some history, perhaps it is
time to plough some of the experience gained back into a new
approach, laying the groundwork for our theory in such a way as to
avoid well-known problems.  The causal set program
\cite{Bombelli:1987aa,Sorkin:1990bh,Sorkin:1990bj,Sorkin:1997,Sorkin:2003bx}
represents such an attempt.

In this review, some answers to the above questions, as embodied
by the causal set program, are set out and explained, and some of
their consequences are given.  As part of this, the results and
open problems in the program are discussed.  In section
\ref{s:motivedefs}, reasons for hypothesising spacetime
discreteness are reviewed.  The definition of a causal set is
given, along with the proposed correspondence principle between
this structure and the effective continuum description of
spacetime.  Then some of the unique features of this
discretisation scheme are discussed.  In section \ref{s:dynamics},
ideas for causal set dynamics are given.  Next, a review is made
of some phenomenological models based on this quantum gravity
program, and successes and challenges in this line of work are
summarised. The final section deals with the issue of recovering
locality for causal sets, something which touches all the other
subjects covered, and new results, which solve this problem in
some situations for the first time, are examined.

Besides the present work, there are many other reviews available.
One of the most recent is \cite{Sorkin:2003bx}, while motivation
and earlier work is covered in
\cite{Sorkin:1990bh,Sorkin:1990bj,Sorkin:1997}.  A philosophically
oriented account of the conception of the causal set idea is given
in \cite{Sorkin:1989re}, and there is a recent review which
introduces some of the core concepts of causal set kinematics and
dynamics \cite{Dowker:2005tz}.  Many of these articles, and other
causal set resources, are most easily found at Rafael Sorkin's web
site \cite{SorkinWeb}.
\section{The causal set approach}
\label{s:motivedefs}

This program is a development of ``path-integral'' or
sum-over-histories (SOH) type approaches (for reasons to adopt
this framework in quantum gravity, see
\cite{Hawking:1980gf,Sorkin:1997gi}).  In such approaches, a space
of histories is given, and an amplitude (or more generally a
quantum measure \cite{Sorkin:1994dt}) is assigned to sets of these
histories, defining a quantum theory in analogy with Feynman's
path integrals.  A basic question, then, is what the space of
histories should be for quantum gravity.  Should they be the
continuous Lorentzian manifolds of general relativity -- or some
discrete structure to which the manifold is only an approximation?

\subsection{Arguments for spacetime discreteness}

A number of clues from our present theories of physics point
towards discreteness. The problematic infinities of general
relativity and quantum field theory are caused by the lack of a
short distance cut-off in degrees of freedom; although the
renormalisation procedure ameliorates these problems in QFT, they
return in naive attempts to quantise gravity (see
\cite{Kalmykov:1995fd} and references therein).  Secondly,
technical problems arise in the definition of a path-integral on a
continuous history space that have never been fully resolved.  On
top of this, the history space of Lorentzian manifolds presents
special problems of its own \cite{CorreiadaSilva:1999cg}. A
discrete history space provides a well defined path integral, or
rather a sum, that avoids these problems.

Perhaps the most persuasive argument comes from the finiteness of
black hole entropy.  With no short-distance cut-off, the so called
``entanglement entropy'' of quantum fields (the entropy obtained
when field values inside a horizon are traced out) seems to be
infinite (see \cite{Bombelli:1986rw,Sorkin:1997ja}, and
\cite{Bekenstein:1994bc,Jacobson:2003wv} for some debate).  If
this entropy is indeed included in the black hole entropy, as many
expect, a short distance cut-off of order the Planck scale must be
introduced to allow agreement with the well-known semiclassical
results.  This, and similar analysis of the shape degrees of
freedom of the black hole horizon
\cite{Sorkin:1995ej,Sorkin:1996sr} lead to the conclusion that
Planck scale discreteness is unavoidable, if the Area-Entropy
relation for black holes is to arise from the statistical
mechanics of a quantum theory.

Finally, suggestions of discreteness have come from various
quantum gravity programs, like loop quantum gravity (see \eg
\cite{Ashtekar:2004eh} and section \ref{s:other} of this article).
Some of the most intriguing results come from so-called ``analogue
models'' \cite{Barcelo:2005fc}, where objects similar to black
holes can be mocked up in condensed state matter systems.  These
analogies, as well as more direct arguments
\cite{Jacobson:1995ab}, suggest that the Einstein equation arises
only as an equation of state, a thermodynamics of some more
fundamental underlying theory.  And the atomic discreteness of
such systems provides a necessary cut-off to degrees of freedom at
small scales.  It is worth considering that atomic discreteness
could never be found, for instance, by quantising some effective
continuum theory describing a gas; it must be an independent
hypothesis.

Quite apart from these more physical arguments, introducing
discreteness can be of great utility.  Conceptual problems, hidden
under layers of technical complexity in continuum treatments, can
sometimes be expressed more clearly in a discrete setting, and
wrestled with more directly.  This quality of discrete models has
been of use in many quantum gravity programs.  The successful
definition of the ``observables'' in the ``classical sequential
growth'' dynamics \cite{Brightwell:2002vw} (see section
\ref{s:CSG}), an analogue of the problem of time in causal set
theory, is an example of this.

\subsection{What kind of discreteness?}
\label{s::defs}

Given these reasons for spacetime discreteness, in what way should
we proceed?  One might be disheartened by the sea of
possibilities; how can we know, at this stage of knowledge, what
the structure underlying the continuum manifold could be? However,
the causal set offers a choice for the histories with a number of
compelling and unique advantages.

The inspiration for the causal set idea comes from the remarkable
amount of information stored in the causal order of spacetime.  It
has been proven that, given only this order information on the
points, and volume information, it is possible to find the
dimension, topology, differential structure, and metric of the
original manifold \cite{Hawking:1976fe,Malament:1977}.  The points
of a (weakly causal \footnote{A weakly causal Lorentzian manifold
is one that contains no closed causal curves, otherwise called
``causal loops''.} ) Lorentzian manifold, together with the causal
relation on them, form a partially ordered set or \textit{poset},
meaning that the set of points $C$ and the order $\prec$ on them
obey the following axioms:

\paragraph{} $(i)$ Transitivity:
  $(\forall x,y,z\in C)(x\prec y\prec z\implies x\prec z).$
\paragraph{} $(ii)$ Irreflexivity:
  $(\forall x\in C)(x\not\prec x).$

\paragraph{}If $x\prec y$ then we say ``$x$ is to the past of $y$'', and if two points of the set $C$ are unrelated by $\prec$ we say they are spacelike (in short, all the normal ``causal'' nomenclature is used for the partial order).

It is this partial order that we choose as fundamental.  To achieve discreteness, the following axiom is introduced:

\paragraph{} $(iii)$ Local finiteness:
  $(\forall x,z\in C) \, ( {\bf card} \SetOf{y\in C}{x\prec y\prec z} < \infty ).$

\begin{figure}[ht]
\centering \resizebox{1.8in}{1in}{\includegraphics{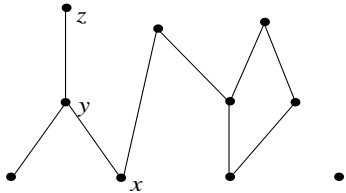}}
\caption{\small{ A causal set. The figure shows an example of a
Hasse diagram.  In such a diagram, the elements of a causal set
are represented by dots, and the relations not implied by
transitivity are drawn in as lines (for instance, because $x\prec
y$ and $y \prec z$, there is no need to draw a line from $x$ to
$z$, since that relation is implied by the other two).  The
element at the bottom of the line is to the past of the one at the
top of the line. }\label{f:hasse}}
\end{figure}

\paragraph{} Where ${\bf card} \, X$ is the cardinality of the set $X$.  In other words, we have required that there only be a finite number of elements causally between any two elements in the structure (the term ``element'' replaces ``point'' in the discrete case).  A locally finite partial order is called a causal set or \textit{causet}, an example of which is illustrated in figure \ref{f:hasse}.  Many researchers have independently been led to the same hypothesis \cite{Myrheim:1978ce,'tHooft:1978id,Bombelli:1987aa}: that the causal set should be the structure that replaces the continuum manifold.

\subsection{The Continuum Approximation}
\label{s:sprinkling}

In all standard quantum theories, be they direct quantisations of
a classical theory or discrete approximations, there is an
approximate correspondence between at least some of the underlying
histories and those of the limiting classical theory, needed in
order to relate the quantum theory to known physics.  (Similarly,
in the state vector formulation, there must be a correspondence
between configurations at a particular time, each assigned a basis
vector in the Hilbert space of the quantum theory, and the allowed
configurations in the classical theory -- something thought to be
true even in loop quantum gravity \cite{Bombelli:2004si}).  The
view taken in causal set theory is that such a correspondence is
necessary for any quantum theory, and so at least some of the
histories in our quantum gravity SOH must be well approximated by
Lorentzian manifolds.  A full justification of this point is
beyond the scope of this article, but support may be taken from
other quantum theories, both standard and speculative, and from
some of the seminal writings on quantum mechanics
\cite{Feynman:1948ur,VonNeumann:1955vi}.  Some relevant quotes
from these and brief explanation will be presented in
\cite{Henson:2005b}.

Here, the question is: when can a Lorentzian manifold $(\cM,g)$ be said to be an approximation to a causet $\cC$?  Roughly, the order corresponds to the causal order of spacetime, while the volume of a region corresponds to the number of elements representing it
\footnote{While this is the stance taken in what might be called the ``causal set quantum gravity program'', the causal set structure has also been useful elsewhere, although with different, or undefined, attitudes as to how it corresponds to the continuum. See for example \cite{Markopoulou:1997wi,Markopoulou:1997hu,Hawkins:2003vc,Foster:2004yc}.}
.  It is interesting to note that the manifold and the metric on it have been unified into one structure, with counting replacing the volume measure; this is a realisation of Riemann's ideas on ``discrete manifolds'' \cite{Riemann} (see also the translated passages in \cite{Sorkin:2003bx}).  But a more exact definition of the approximation is needed.

\begin{figure}[ht]
\centering
\resizebox{1in}{0.6in}{\includegraphics{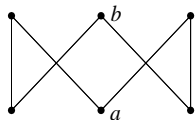}}
\caption{\small{
A Hasse diagram of the ``crown'' causet.  This causet cannot be embedded in 1+1D Minkowski space: if the above Hasse diagram is imagined as embedded into a 2D Minkowski spacetime diagram, the points at which elements $a$ and $b$ are embedded are not correctly related.  In no such embedding can the embedded elements have the causal relations of the crown causet induced on them by the causal order of 1+1D Minkowski space.  The causal set can however be embedded into 2+1D Minkowski space, where it resembles a 3 pointed crown, hence its name.
}\label{f:crown}}
\end{figure}

A causal set $\cC$ whose elements are points in a spacetime $(\cM,g)$, and whose order is the one induced on those points by the causal order of that spacetime, is said to be an \textit{embedding} of $\cC$ into $(\cM,g)$
\footnote{Really an embedding of the isomorphism class of that causet (the ``abstract causet'').  The distinction between isomorphism classes and particular instances of causal sets is not crucial for the purposes of this article, and will be ignored.}
.  Not all causal sets can be embedded into all manifolds.  For example, the causal set in figure \ref{f:crown} cannot be embedded into 1+1D Minkowski space, but it \textit{can} be embedded into 2+1D Minkowski space.  There are analogues to this causal set for all higher dimensions \cite{Brightwell:1989}, and surprisingly there are some causal sets that will not embed into Minkowski of \textit{any} finite dimension. Thus, given a causal set, we gain some information about the manifolds into which it could be embedded.  However, a manifold cannot be an approximation to every causal set that embeds into it;  we could recover no volume information in this way, no discreteness scale is set, and there might not be enough embedded elements to ``see'' enough causal information.  A further criterion is needed to ensure the necessary density of embedded elements.

So, to retrieve enough causal information, and to add the volume information, the concept of \textit{sprinkling} is needed.  A sprinkling is a random selection of points from a spacetime according to a Poisson process.  The probability for sprinkling $n$ elements into a region of volume $V$ is

\begin{equation}
\label{e::poisson}
P(n)= \frac{(\rho V)^n e^{-\rho V}}{n!}.
\end{equation}

Here, $\rho$ is a fundamental density assumed to be of Planckian order.  Note that the probability depends on nothing but the volume of the region.  The sprinkling defines an embedded causal set.  The Lorentzian manifold $(\cM,g)$ is said to approximate a causet $\cC$ \textit{if $\cC$ could have come from sprinkling $(\cM,g)$ with relatively high probability}
\footnote{The practical meaning of ``relatively high probability'' has so far been decided on a case-by-case basis. It is usually assumed that the random variable (function of the sprinkling) in question will not be wildly far from its mean in a faithfully embeddable causet.  Beyond this, standard techniques involving $\chi^2$ tests exist to test the distribution of sprinkled points for Poisson statistics.}
.  In this case $\cC$ is said to be \textit{faithfully embeddable} in $\cM$.  On average, $\rho V$ elements are sprinkled into a region of volume $V$, and fluctuations in the number are typically of order $\sqrt{\rho V}$ (a standard result from the Poisson statistics), becoming insignificant for large $V$.  This gives the promised link between volume and number of elements.

Can such a structure really contain enough information to provide a good manifold approximation?  We do not want one causal set to be well-approximated by two spacetimes that are not similar on large scales.  The conjecture that this cannot happen (sometimes called the ``causal set haupvermutung'', meaning ``fundamental conjecture'') is central to the program.  It is proved in the limiting case where $\rho \rightarrow \infty$ \cite{Bombelli:1989mu}, and there are arguments and examples to support it, but some steps remain to be taken for a general proof.  One of the chief difficulties has been the lack of a notion of similarity between Lorentzian manifolds, or more properly, a distance measure on the space of such manifolds.  Progress on this has now been made \cite{Noldus:2003si}, raising hopes of a proof of the long-standing conjecture.

A further generalisation of this scheme may be necessary.  Above, it was noted that certain small causal sets cannot be embedded into Minkowski space of any particular dimension.  This means that, for $\cC$ a large causal set that is faithfully embeddable into a region of $n$-dimensional Minkowski, by changing a small number of causal relations in $\cC$ we can form a causet that no longer embeds.  From our experience with quantum theories, we most likely will not want such ``small fluctuations'' to be physically significant, and so we may need a condition of ``manifoldlikeness'' that is more forgiving than faithful embedding.  A possible method is given by \textit{coarse-graining} \cite{Sorkin:1990bh}: removal of some points from the causal set $\cC$ forming a new causal set $\cC'$, before testing $\cC'$ for faithful embeddability at the appropriate lower density of sprinkling $\rho'$.  For example, this might reasonably be done at random with the same probability $p$ for removal of each element, and $\rho'=\rho (1-p)$.  This basically amounts to looking for a faithfully embeddable subset of a causal set, following a certain set of rules.  Below, the criterion of faithful embeddability will be the one used, but it should be kept in mind that the causets being talked about could be coarse-grainings of some larger causet.

\subsection{Reconstructing the continuum}

The concept of faithful embedding gives the criterion for a
manifold to approximate to a causet.  But it is important to
realise that the only use of sprinkling is to assign continuum
approximations; the causal set itself is the fundamental
structure.  How then can this approximate discrete/continuum
correspondence be used?

An array of techniques exist to recover approximate manifold
information from a given causal set.  For example, functions
$D(\cC)$ of the causal set structure can be found which, for $\cC$
a random sprinkling of $\cM$, accurately approximate the dimension
of $\cM$ with very high probability
\cite{Meyer:1988,Myrheim:1978ce,Reid:2002sj}.  Knowing this, given
a (suitably large) causet $\cC$ we can say ``If this causet
corresponds to a manifold, then its dimension is near $D(\cC)$''.
Conversely, if two such dimension estimators do not match on large
scales, we can be sure that $\cC$ does not correspond to a
continuum spacetime at all.  Such estimators exist for timelike
distance between elements \cite{Brightwell:1990ha,Ilie:2005qg},
and of course volumes.  As another example, methods have been
developed to retrieve topological information about spatial
hypersurfaces in approximating Lorentzian manifolds, by reference
only to the underlying causet \cite{Major:2006}.

A simple example of how such estimators work is given by one of the estimators of timelike distance.  In Minkowski space, the timelike distance between two points is related to the volume of the interval between them (the interval is the region that is causally between the two points).  In four dimensions the relationship is
\begin{equation}
\label{e::dist}
Vol \Bigl( I(x,y) \Bigr)=\frac{\pi}{24} d(x,y)^4,
\end{equation}
where $x$ and $y$ are the timelike related points, $d(x,y)$ is the distance between them, $Vol(R)$ is the volume of region $R$ and $I(x,y)$ is the interval between $x$ and $y$.  Directly from the concept of faithful embedding, we know that the number of elements sprinkled into a region is a good approximation to the volume of that region in fundamental units, and that there are only small proportional fluctuations for large volumes.  Therefore, given two timelike related elements in a causet, the volume of the interval between them can be estimated, and equation (\ref{e::dist}) (or its analogue in different dimensions) used to find the distance between them.  See \cite{Brightwell:1990ha} for a different distance measure conjectured to hold for curved spacetimes \cite{Myrheim:1978ce}, and a way to identify approximations to timelike geodesics.  As this other distance measure does not depend on the dimension (up to a multiplicative constant), the two can be compared to give one of the dimension estimators mentioned above.

Given a causal set $\cC$ without an embedding (this is after all
our fundamental structure) it would be of great utility to be able
to say if it was faithfully embeddable into some spacetime or not
-- a criterion of ``manifoldlikeness'' -- and if so to provide the
embedding.  The discrete-continuum correspondence given above does
not directly answer this question; it would be highly impractical
to carry out various sprinklings until we came up with a causet
isomorphic to $\cC$.  Nevertheless, the measure of timelike
distance and some simple geometry can be used, on computer, to
attempt to find an embedding for $\cC$, at least into small
regions of Minkowski \cite{Henson:2006dk}.  The idea is applicable
to causets that are embeddable into manifolds of any dimension,
but has, so far, been implemented only in 2D.  The success or
failure of the attempted embedding gives a measure of
manifoldlikeness for $\cC$.   Similar embedding methods have also
been investigated by David Reid.  Beyond this rough-and-ready
computational scheme, several necessary conditions for
manifoldlikeness are known (\eg the matching of different
dimension estimators, ``self-similarity'' \cite{Rideout:2000fh},
\etc), and it is hoped that a combination of these might yield a
necessary and sufficient condition.

Given this discrete/continuum correspondence principle, some of
the attractive features of the causal set structure can be noted.
Firstly, there is no barrier to sprinkling into manifolds with
spatial topology change, as long as it is degeneracy of the metric
at a set of isolated points that enables topology change, and not
the existence of closed timelike curves (one of these conditions
must exist for topology change to occur, see \eg
\cite{Borde:1999md} and references therein) -- and in this
discrete theory there is no problem with characterising the set of
histories.  For those who believe that topology change will be
necessary in quantum gravity \cite{Sorkin:1997gi,Hawking:1978zw},
this is important.  Secondly, the structure can represent
manifolds of any dimension -- no dimension is introduced at the
kinematical level, as it is in Regge-type triangulations.  In
fact, scale dependent dimension and topology can be introduced
with the help of course-graining, as explained in
\cite{Meyer:1988}, giving an easy way to deal with notions of
``spacetime foam''.  Also, it has been found necessary to
incorporate some notion of causality at the fundamental level in
other approaches to SOH quantum gravity, highlighting another
advantage of using the causal set structure from the outset.  But
the property which really sets causal sets apart from other
discrete structures is local Lorentz invariance.

\subsection{Lorentz Invariance and discreteness}

For most discrete structures, local Lorentz invariance (LLI) is
impossible to attain (see \cite{Dowker:2003hb} for a brief
explanation of why this is so).  This can be a major problem if
the locally Lorentz invariant spacetime we observe is to arise as
an approximation to these structures.  There is always the
possibility that LLI does fail at higher energy scales, and
discreteness of the Lorentz violating kind has been cited as a
motivation when searching for such non-standard effects.  As such
studies progress, bounds on Lorentz violation from astrophysical
observations are becoming ever more stringent
\cite{Jacobson:2005bg,Amelino-Camelia:2004ht}.  On top of this,
Collins \textit{et al.} \cite{Collins:2004bp,Perez:2003un} argue
that Lorentz symmetry breaking at the Planck scale would
significantly affect the radiative corrections in the standard
model.  This means that, at best, additional fine-tuning would
have to be introduced to maintain agreement with experiment.

What does Lorentz invariance mean in this context?  The standard
concept only makes sense at the level of the continuum (and global
Lorentz invariance is present only in Minkowski space).  So, it is
at the level of the continuum approximation that we must think
about LLI.  Consider a discrete structure that has Minkowski space
as an approximation.  If the underlying structure, in and of
itself, serves to pick out a preferred direction in the Minkowski
space, then Lorentz invariance has been violated.  This is the
situation for lattice-like structures.

In contrast, the causal set provides a locally Lorentz invariant discrete structure -- the only one considered in any approach to quantum gravity.  This property is achieved thanks to the random nature of the discrete/continuum correspondence principle given above.

In analogy, consider a crystal, and a gas, as discrete systems of atoms whose behaviour can be given an approximate continuum treatment.  The crystal has a regular underlying structure that breaks rotational symmetry, and this symmetry breaking can be observed macroscopically, by the existence of fracture planes and so on.  The gas on the other hand has a \textit{random} underlying structure, and the probability distribution of the molecules' positions at any time is rotationally invariant.  There is no preferred direction in a gas that affects its behaviour in the effective continuum treatment.  We could ``cook up'' a direction from the positions of the molecules -- in any region containing two molecules we can of course draw a vector from one to the other.  The point is that  such ``preferred directions'' identifiable on microscopic scales have no effect on the bulk, continuum physics of the gas.  Thus it is common to say that the behaviour of a gas is rotationally invariant.

The Lorentz invariance of the causal set is similar.  As previously noted, in the Poisson process, the probability for sprinkling $n$ elements into a region depends on no property of that region other than the volume.  In Minkowski spacetime, to establish Lorentz invariance of the Poisson process rigorously we need only note the theorems proving the existence and
uniqueness of the process with the distribution (\ref{e::poisson}) for all
measurable subsets of $\Reals^d$ and its invariance under all
volume preserving linear maps (see \eg \cite{Stoyan:1995}), which of course includes Lorentz transformations.  As with the gas, any particular instance of the process -- a particular sprinkling -- will have directions identifiable in small regions containing few points, but this will have no effect on the continuum treatment. In a curved
spacetime, Lorentz invariance is to be understood to hold in the
same sense that it holds in General Relativity: the equivalence of
local Lorentz frames.

In some sense the situation is better than that for gases.  In the case of a sprinkling of $\Reals^3$, a direction can be associated with a point in the sprinkling, in a way that commutes with rotations (\ie, finding the direction from the sprinkling and then rotating the direction gives the same result as first rotating the sprinkling and then finding the direction).  An example is the map from the point in the sprinkling to the direction to its nearest neighbour.  But due to the non-compactness of the Lorentz group, there is no way to associate a preferred frame to a point in a sprinkling of Minkowski that commutes with Lorentz boosts
\footnote{A theorem expressing this will be exhibited in future work of Luca Bombelli, Rafael Sorkin and the author.}
.

For causal sets that approximate to Minkowski space, the causal set does not pick out a preferred direction in the approximating manifold.  We therefore expect that no alteration to the energy-momentum dispersion relation would be necessary for a wave moving on a causal set background.  This property is explicit in at least one simple model \cite{Dowker:2005}.

Local Lorentz invariance of the causal set is one of the main things that distinguishes it among possible discretisations of Lorentzian manifolds.  We now see that the daunting choice of the discrete structure to be used in quantum gravity is actually extremely limited, if the principle of local Lorentz invariance is to be upheld.  Could other popular approaches to quantum gravity, based on graphs and triangulations, utilise sprinklings to incorporate Lorentz symmetry?  The theorem mentioned above shows this to be impossible: if no direction can be associated to a sprinkling of Minkowski in a way consistent with Lorentz invariance, how could an entire finite valancy graph or triangulation?

\subsection{LLI and Discreteness in other approaches}
\label{s:other}

In the causal set approach, there is discreteness and LLI at the
level of the individual histories of the theory.  This seems to be
the most obvious way to incorporate the symmetry, while ensuring
that the foreseen problems with black hole entropy (and other
infinities) are avoided.  But is it necessary?  Each approach to
non-perturbative quantum gravity represents a different view on
this, some of which can be found in the other chapters of the book
for which this article was prepared \cite{Oriti:2006}. A brief
``causal set perspective'' on some of these ideas is given here.

In the broad category of ``spin-foam approaches'', the histories are also discrete, in the sense that they can be seen as collections of discrete pieces of data.  But is there ``enough discreteness'' to evade the infinite black hole entropy arguments?  This is a hard question to answer, not least because the approximate correspondence between these histories and Lorentzian manifolds has not been made explicit
\footnote{However, the correspondence principle in the similar case of graphs corresponding to 3D space \cite{Bombelli:2004si} could possibly be extended, somehow, to the 4D case.}
.  But there is intended to be a correspondence between the area of a 2D surface and the number (and labels) of 2-surfaces in the spin-foam that ``puncture'' it, suggesting a kind of fundamental discreteness on such surfaces.  It also suggests an upper bound on degrees of freedom per unit volume.  But all this depends on the final form of the sum over triangulations in that approach, something not yet clarified.

The status of Local Lorentz invariance in spin-foam models remains controversial.  If fundamental discreteness \textit{is} maintained in the spin-foam path integral (\ie if spin-foams with Planck-scale discreteness dominate semi-classical states), there is good reason to conclude that problems with LLI will result.  Spin-foams with Planck scale discreteness, like any lattice-like structure, are not Lorentz invariant:  near one particular frame, a good approximation to the large-scale properties of a ``nicely shaped'' continuum region may be achieved, but away from that frame a region of the same shape will be badly approximated.  It is sometimes claimed that, although an individual spin-foam does not satisfy LLI, a quantum sum over many spin-foams may do (arguments from the closely related LQG program support this \cite{Rovelli:2002vp,Livine:2004xy}).  An analogy is drawn with rotational invariance: in that case, the histories might only represent one component of the angular momentum of, say, an electron.  In spite of this, the physics represented is in fact rotationally invariant.  However, in \cite{Henson:2005b} it is argued that this analogy will fail: if spin-foam histories are used, then even \textit{macroscopic} properties of the universe will be not be represented (to any acceptable level of accuracy) in some frames, and this must lead to Lorentz violation.  It is one thing for some of the components of the angular momentum of an electron to be immeasurable, but reasonable approximations to the angular momentum components of a baseball are quite another matter.  Analogous quantities could fail to be measurable in the case of Lorentz transformations of spin-foams.  It is possible that further thought along these lines could lead to quantitative predictions of Lorentz violation from spin-foam models, giving an opportunity for observational support or falsification.

A number of assumptions must be made to reach this conclusion, principally that there would indeed be fundamental discreteness.  But whether or not this is true, and whatever is observed, there is an argument that at least some of the fundamental histories of a theory must be locally Lorentz invariant (or at least give equally good approximations to the continuum in all local frames), if locally Lorentz invariant phenomenology is to be predicted.

In the loop quantum gravity program, the spectra of certain operators (\eg the areas of 2D surfaces) are claimed to be discrete, although as yet the physical Hilbert space and operators have not been identified.  Nevertheless, some arguments have been provided as to how the problems of spacetime singularities and black hole entropy might be solved in LQG.  But without the physical observables, how this type of discreteness could circumvent the arguments mentioned in the introduction or in the previous paragraph, or even whether it would exist in a completed form of loop quantum gravity, is not clear as yet.

In dynamical triangulations, discreteness is used to solve the problems of defining the path integral, and coming to grips with technical issues in a manageable way, notably the Wick rotation.  However, in this approach the discreteness is not considered fundamental and a continuum limit is sought.  As the ``causal dynamical triangulations'' program is in the happy situation of possessing a working model, it would be of great interest for the debate on discreteness to see what becomes of black hole entropy (or more general forms of horizon entropy \cite{Jacobson:2003wv}) as the cut-off is removed.  Will some previously unexpected effect keep the entropy finite, or are the arguments for a fundamental cut-off inescapable?

\section{Causal Set Dynamics}
\label{s:dynamics}

Having introduced the causal set structure and discussed some of
its special advantages, the most pressing question is how to
construct a dynamics, with causal sets as the histories, that
would be a satisfying theory of quantum gravity.  The question is,
perhaps unsurprisingly, a difficult one.  In order to obtain the
discrete, Lorentz invariant causal set we have thrown away much of
the manifold structure that we are used to.  For instance, the
idea of geometry on a spacelike slice is only an effective one.
The analogue to a spacelike surface in a causal set $\cC$ is a
\textit{inextendable antichain}: a set of elements $\cA$ of $\cC$
that are all spacelike to each other, and such that no more
elements of $\cC$ can be added to $\cA$ that are spacelike to all
the elements already in $\cA$.  Such a set contains next to no
information:  the only structure is the number of elements in the
antichain.  Only when the antichain is sufficiently ``thickened''
(adding elements close to $\cA$ in its past and/or future), can
more pieces of approximate manifold information be recovered
\cite{Major:2006}.  Because of this, a state-vector approach to
quantum mechanics, with a configuration space representing the
degrees of freedom on a spacelike slice, seems particularly
unsuited to causal sets.  This is not viewed as a drawback, as the
program was motivated by the SOH view of quantum mechanics and
seeks to avoid the problem of time as it is encountered in
canonical approaches.  Also, the universe is expanding, suggesting
that in any approach with an upper bound on degrees of freedom per
unit volume, the Hilbert space describing the universe would have
to grow with time.  This, and the well-known black hole
information loss problems, suggest that a successful quantum
gravity theory will not be unitary (see the chapter by Rafael
Sorkin in \cite{Oriti:2006}).

This begs the question: what other type of dynamics could be used?
Even the Feynman path integral crucially refers to states on
spacelike hypersurfaces.  Is there a dynamical framework that is
truly free of references to states like this?  Such a ``spacetime
approach'' to quantum mechanics is natural for a generally
covariant theory in any case, and one has been developed:
generalised quantum mechanics \cite{Hartle:1992as}, alternatively
named quantum measure theory \cite{Sorkin:1994dt} (an introduction
appears in \cite{Oriti:2006}).  The idea is that quantum processes
can be described as a generalisation of stochastic processes; the
quantum measure, unlike the probability measure, is not
necessarily additive on disjoint sets of histories, allowing for
interference effects. The framework includes standard quantum
mechanics as a special case and also allows non-unitary dynamics.
Most ideas for causal set dynamics are based on this framework.

\subsection{Growth models}
\label{s:CSG}

The most favoured approach to dynamics uses the simple structure and direct physical interpretation of the causal set to advantage.  Given such a simple kinematical framework, and the dynamical framework of quantum measure theory, it is possible that physical principles could be used to constrain the dynamics until only a small class of theories remain (the ideal example being the derivation of GR from a small set of such principles).  Particularly natural to the causal set are the concepts of general covariance and causality. As a warm-up for the quantal case, a set of stochastic processes on the space of past-finite causal sets has been developed, the so-called \textit{classical sequential growth} (CSG) models\cite{Rideout:1999ub,Varadarajan:2005gg}.

\begin{figure}[ht]
\centering
\resizebox{5.1in}{3.0in}{\includegraphics{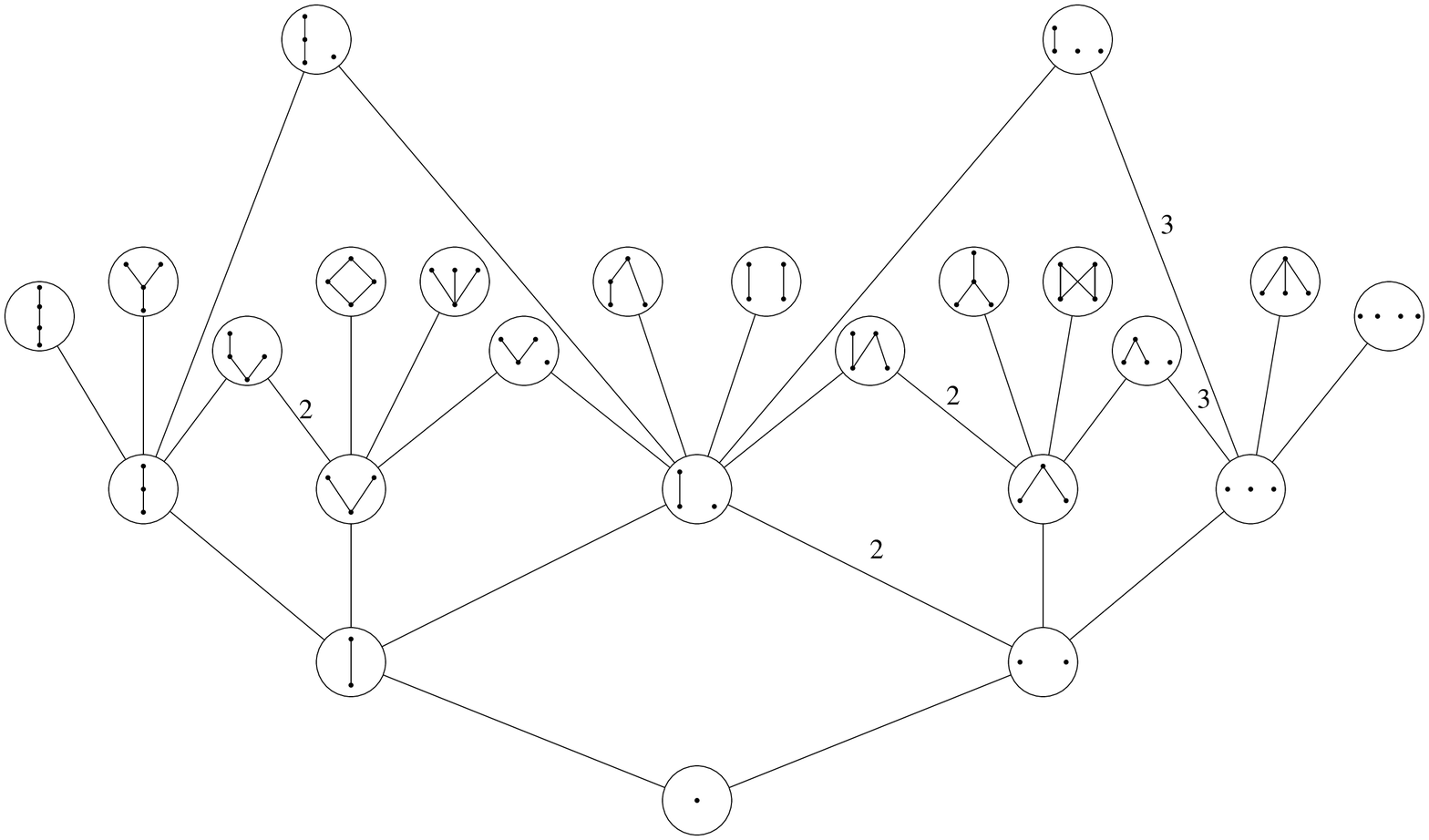}}
\caption{\small{
An augmented Hasse diagram of ``poscau'', the partially ordered set of finite causets.  The elements of this set are the finite causets.  To the ``future'' of each causet are all the causets that can be generated from it by adding elements to the future of or spacelike to its elements (the numbers on some of the links represent the number of different ways the new element can be added, due to automorphisms of the ``parent'' causet).  Only causets of up to size 4 are shown here.  An upwards path in poscau represents a sequence of transitions in a growth process.  Each such path is given a probability by a CSG model.  Because of the general covariance condition, the probabilities of paths ending at the same causet are the same.  (Note that the apparent ``left-right symmetry'' of poscau does not survive above the 4-element causets.)
}\label{f:poscau}}
\end{figure}

These models are based on the concept of randomly ``growing'' a causal set, following certain rules. From a single element, one new element is added in each of an infinite sequence of \textit{transitions}.  The new element is always added to the future of or spacelike to the existing elements.  There are always several possibilities for how to add the new element, and a probability distribution is placed on these possibilities.  These ``transition probabilities'' are then constrained by the chosen physical principles.  A sequence of transitions can be thought of as a path through the partially ordered set of all finite causets, as illustrated in figure \ref{f:poscau}.  The process generates infinite-element causal sets.  From the transition probabilities, a probability measure on the space of all infinite-element past-finite causal sets can be constructed.

The order of birth can be viewed as a labelling of the elements of the growing causet.  A natural implementation of the principle of general covariance is that this labelling should not be significant for the dynamics.  More exactly, the probability of reaching a certain finite causet in the growth process should not depend on the order in which the elements were added.  Another physical principle is introduced to ban superluminal influence.  This is done in a way appropriate to stochastic systems, by a causal condition similar to that used by Bell to derive his famous inequalities.  With these constraints, the free parameters of the model are reduced to one series of real numbers.

The CSG models have made a useful testing ground for causal set dynamics, allowing some questions to be answered that would have relevance for quantum theories developed using the same method.  For instance, the ``typical'' large causal set (\ie the type that is most likely to be found from a uniform probability distribution over causal sets with some large number of elements) does not look like a manifold, but instead has a ``flat'' shape described more fully in \cite{Kleitman:1975}.  It might be wondered what kind of a dynamics could overcome the great numbers of these ``Kleitman-Rothschild'' causets -- entropic effects might be expected to favour these typical causets.  However, growth process models easily circumvent this worry.  CSG models generically give low probabilities to these causets.

An important question is how to identify and characterise the physical questions that the theory can answer.  In canonical theories this question can be phrased ``what are the physical observables, and what do they mean?'', and answering it is a central part of the problem of time.  In some form it seems to afflict any indeterministic, generally covariant theory.  This problem does not disappear in the CSG model, where it must be ensured that the ``observables'' (here, sets of histories which are assigned a probability, and are ``covariant'', meaning insensitive to the growth order labelling) can be characterised and given a physical interpretation.  This was achieved for a generic class of CSG models in \cite{Brightwell:2002vw} and the results extended to the most general models in \cite{Dowker:2005gj}.  Most of the methods used will be directly applicable to any future ``quantum sequential growth'' model.

Another result with possible implications for the full quantum theory concerns the so-called ``cosmic renormalisation'' behaviour that the models exhibit \cite{Sorkin:1998hi,Martin:2000js,Ash:2002un}.  Certain generic CSG models yield a ``bouncing'' cosmology, in which the universe expands from zero size to a certain ``spatial volume'' (measured by the size of maximal antichains), and then contracts down to minimal size, only to expand once again, in an infinite chain of big bang/big crunch cycles.  The dynamics after each new big bang is equivalent to a CSG model with no previous cycles, but with a different ``renormalised'' set of free parameters.  After a large number of cycles the values of these effective parameters converge to a small set of those possible -- in particular, the total size of the cycle (\ie the number of elements between its big bang and crunch) becomes large whatever the size in the first few cycles.  This gives a possible mechanism for the setting of fundamental constants: the large spatial extent of the universe in these models is not a result of fine-tuning, but simply a consequence of the extreme age of the universe.

The CSG models have also been of use in developing tests of dynamically generated causal sets to look for manifoldlike behaviour \cite{Rideout:2000fh}, and computational techniques for causal sets.  But it is important to note that the theories are not supposed to be a ``classical limit'' of a quantum dynamics;  the situation is more analogous to the stochastic dynamics of Brownian motion, and its relationship with the quantum dynamics of the Schr\"odinger particle.  The goal is to replace the probability measure used in the CSG model with a quantum measure, reworking the physical conditions to make sense in this case.  Whether or not the CSG models can produce manifoldlike causal sets is not crucial for them to fulfil their role as a stepping-stone to the quantum case.

This ambitious approach to causal set dynamics has the advantage of simple, clean formalism and the prospect of going beyond what might be possible by attempting to approximate a continuum path integral.  For instance, no dimension is specified anywhere in the founding principles of the theory, and so a successful ``quantum sequential growth'' model would give a real explanation for the 4D nature of large-scale spacetime from a small set of principles.  However, challenges remain in the development of the full quantum theory.  The generalisation of the relativistic causality principle to the quantum case has proved difficult \cite{Henson:2005wb}.  It must also be ensured that there is as little freedom in the implementation of the fundamental principles as possible, lest we undermine the idea of directly proceeding from principles to dynamics.  This question deserves a more thorough investigation even at the stochastic level.  But many avenues for creating growth models lie open.  Different physical principles could be used to constrain the dynamics, and a number of suggestions are currently under consideration.

\subsection{Actions and Amplitudes}
\label{s:actions}

Another approach to the dynamics is in closer analogy to that employed in other quantum gravity programs: assigning a complex amplitude to each history.
  It would be interesting to see how far this formalism can be pushed for causal set dynamics (its use was originally suggested in the early papers in the program \cite{Bombelli:1987aa,Sorkin:1990bh}).  The first obstacle is the lack of an expression for the amplitude $\exp{(i\,S(\cC))}$.  We need to find an action for the causal set.  The most obvious thing to do would be to find a function of the causal set that approximates the Einstein action for causal sets corresponding to 4D manifolds.  This is another kinematical question like that of finding geodesic lengths, and dimension, discussed above.  In the continuum, the Einstein action is the integral of a local quantity on the manifold, and so the causal set action should also be local, approximately.  Indeed, the ``natural'' value of any such approximately local function should approximate the Einstein action, as argued in \cite{Bombelli:1987aa}.  Then the kinematical task becomes the identification of approximately local causal set functions.  The task of recovering locality has been a perennial theme of research in causal set theory, and has recently seen some exciting progress, leading to some possible expressions for the action, as discussed in section \ref{s:locality}.

Then there is the question of what set of causal sets to sum over.
Most satisfying would be to sum over \textit{all} causal sets of a
fixed number of elements (a ``unimodular'' sum over histories
\cite{Sorkin:1987cd,Unruh:1988in,Sorkin:1988a}).  The action would
have to be ``slowly varying'' in some appropriate sense near the
classical solutions, and ``quickly varying'' elsewhere -- here
``elsewhere'' means not only the causal sets corresponding to
manifolds that are not solutions of GR, but also the (far greater
number of) causal sets that do not correspond to \textit{any}
manifold.  A less natural, but perhaps useful, strategy would be
to limit the history space to those causal sets that are
faithfully embeddable into certain manifolds.  To make a model
that effectively gives 2D quantum gravity would seem especially
simple using this technique, as here the action is trivial for
manifolds of fixed topological genus, (and in that case the only
degree of freedom on a spatial slice is its length, which
\textit{is} easily recoverable from the causal set, as it is
approximated by the number of elements in the corresponding
inextendable antichain).  Also, the only degrees of freedom on a
2D Lorentzian manifold can be given in terms of the volume
measure.  Varying the conformal factor can be seen as varying the
density of sprinkling for the causal set, and so the history space
can be taken to be the set of all causets embeddable (but not
necessarily faithfully embeddable) into a fixed Lorentzian
manifold of, say, topology $(S^1 \times \Reals)$.  Some
mathematical results about causal sets embeddable into 2D
Minkowski are already known. Interestingly, a typical large causal
set of this type can be shown to be \textit{faithfully} embeddable
into a causal interval of Minkowski \cite{Winkler:1990}.  The
existence of techniques to prove such statements suggests that
analytical results will be possible in the 2D dynamics.  Thus some
of the prerequisites are fulfilled for using the causal set
structure as the regulator for 2D quantum gravity, creating a
model similar to those using dynamical triangulations as
histories.  This makes an interesting project for the near future.

\section{Causal set Phenomenology}

While progress is being made on the dynamics, a final theory is still not available.  But it is still of use to ask the question ``What are the consequences of the causal set hypothesis for phenomenology?''  Does the use of Lorentz invariant, discrete histories suggest any measurable effects?  Without the final dynamics any arguments will have to be heuristic; but, when it comes to phenomenology, that is perfectly good practice, as noted in \cite{Amelino-Camelia:2004yd}.  Such considerations have led in many interesting directions.  One prediction is that no violation of standard, undeformed LLI (as opposed to the deformed Lorentz invariance of ``doubly special relativity'', based on structures other than Minkowski space) will be observed, as such an observation would undermine one of the major motivations for causal set theory.  But this is a purely negative prediction, so it is useful to search for something more.

\subsection{Predicting $\Lambda$}

Perhaps the most significant phenomenological result for causal sets was the successful prediction of the cosmological constant from a heuristic argument \cite{Sorkin:1990bh,Ahmed:2002mj,Sorkin:2005fu}.  The reasoning used is briefly reviewed here.

It is natural in causal set theory to use the number of elements $N$ in the universe ``so far'' as a sort of ``parameter time'' -- this is certainly the case in the CSG models, where the number of elements increases one by one as new elements are added.  Thus, for a quantum theory of causal sets, the sum-over-histories will be over all causal sets of a fixed number of elements, rather than all those with a certain proper-time interval between some initial and final surfaces.  At the level of the continuum, this approximately corresponds to a sum at fixed volume $V$.  This is an acceptable way of performing the gravitational path integral \cite{Sorkin:1987cd,Unruh:1988in,Sorkin:1988a}, described as ``unimodular'', and there is a corresponding unimodular modification of GR which is equivalent to the standard theory.  From the classical theory, it can be seen that the volume $V$ is conjugate to the cosmological constant $\Lambda$, in the sense that position and momentum are conjugate in particle dynamics (this can be seen more or less directly from the way that $\Lambda$ enters into the action for GR, in the term $-\Lambda V$).  So it seems that, if $V$ took an exact value, then the value of $\Lambda$ would be totally undefined, by the uncertainty principle.  How then can a unimodular quantum gravity work?

The answer is that the volume will not be held exactly fixed in our causal set path integral -- this is impossible because the volume is only an approximate concept, related to the number of elements $N$ in the universe.  Crucially, we have a precise idea of how uncertain the volume is for a given value of $N$, given by the discrete/continuum correspondence scheme of section \ref{s:sprinkling}.  The number of elements sprinkled into a given volume undergoes Poisson type fluctuations, giving rise to an uncertainty in the volume of the universe of order $\pm \sqrt{V}$, where $V$ is the past 4-volume of the universe in fundamental units.  Plugging this into the uncertainty relation, we have
\begin{equation}
\Delta \Lambda \sim \frac{1}{\Delta V} \sim \frac{1}{\sqrt{V}},
\end{equation}
using fundamental units.  If we assume that the value of the cosmological constant is driven towards zero (taken as a natural assumption here), this equation tells us that it could not be exactly zero in our theory, but will have fluctuations of order $10^{-120}$ (again in Planckian units) in the present epoch.  This prediction was subsequently verified by observation.

There are plenty of open questions surrounding this achievement.
By this argument, the energy density in $\Lambda$ is, \textit{on
average}, comparable to the matter and radiation energy density at
all times.  However, fluctuations in $\Lambda$ are to be expected,
and in \cite{Ahmed:2002mj} these fluctuations are modelled.  The
hope is that this will lead to more detailed predictions in
cosmology.  This is particularly timely, as observations of Gamma
Ray Bursts are now being used to constrain the past values of the
cosmological constant.  The path from theory to prediction in
cosmology is typically a tortuous one, and the introduction of a
varying cosmological ``constant'' breaks the assumptions used in
standard cosmology. Much effort will be required to modify the
standard predictions in the light of this idea, and then compare
to observation.

\subsection{Swerving particles}
\label{s:swerves}

How might a particle propagate on a causal set?  Is it necessary,
or at least natural in some sense, for them to move in a
non-standard way?  This question has obvious phenomenological
implications, and has been discussed in \cite{Dowker:2003hb}.  In
the continuum, of course, classical particles move on geodesics.
The correct future evolution can be determined by the position and
velocity at the initial time.  But consider a point particle
moving on a causal set background.  Since we want to know what
happens when there is a continuum approximation, we use a causet
from a sprinkling.  How close can we come to geodesic motion?  The
most obvious way to model the path of a point particle is by a
\textit{chain} in the causal set - a subset of elements all of
which are related (this is analogous to a timelike path in the
continuum).  Because of the non-local nature of the causal set,
the velocity of the particle is not exactly defined at each
element.  In order to get an approximation to the velocity, some
section of the chain to the past of the point needs to be
considered.  But we know that physics is approximately local.  A
strict version of this approximate locality is that the dynamics
of the particle to the future of an element is only affected by
the past within some maximum proper time of that element.  If this
is true, then the velocity cannot be obtained to arbitrary
accuracy, and therefore the particle is subject to random (Lorentz
invariant) acceleration -- it ``swerves'' away from the geodesic.
Just as the diffusion equation arises from the standard random
walk, a Lorentz invariant diffusion in velocity space can be found
to model the swerves.  The classical point particle picture, and
the use of a fixed background causal set, are of course gross
simplifications, and the strict form of approximate locality is by
no means an absolute requirement.  But the model is potentially
testable, and that is the goal of studies of this heuristic,
phenomenological type.

One of the mysteries of current astronomical observations is the
origin of high energy cosmic rays (HECRs).  Explanations by known
astronomical sources struggle to reach the gigantic energy scales
required for the highest energy rays, added to which the source of
rays seems to be spread over the whole sky, not at a few localised
places.  The power spectrum of the rays also displays power-law
behaviour across a vast energy range.  Could this be explained by
random acceleration of Hydrogen atoms and other material in the
intergalactic voids?  Applying the velocity diffusion law produced
by the ``swerves'' model,  a power law is indeed obtained
\cite{Dowker:u}. But the rate of diffusion needed to explain
HECR's is incompatible with laboratory requirements.  It is hoped
that this might be corrected in a more sophisticated model that
takes into account the quantum behaviour of the particle and/or
spacetime, since quantum diffusion can produce far different
results to its classical counterpart.  Whatever the outcome, any
model coming from quantum gravity that could solve the HECR
problem would be welcomed by astrophysicists.

\section{The Recovery of Locality}
\label{s:locality}

The issue of approximate locality is a crucial one for many aspects of the causal set program.  It has already been mentioned in the contexts of kinematics, dynamics and phenomenology.  But why should it be difficult to obtain locality at the effective, continuum level?  More specifically, we might ask how to approximate local differential operators on a field.  Let us compare the problem in causal sets to its easy resolution on a light-cone lattice.

The lattice we will consider is a discretisation of 2D Minkowski space.  It is most convenient to use the light-cone co-ordinates $u$ $v$, where the metric is given as $ds^2=2dudv$.  We will consider a scalar field, \ie a function $\phi(u,v):(u,v) \longrightarrow \Reals$.  In the discretisation, lattice sites appear at the points $(na,ma)$ where $a$ is the lattice spacing and $n,m \in \Integers$ where $u$ $v$ are the co-ordinates of a lattice site.  Here, we are helped considerably by the fact that ``nearest neighbours'' to any particular point can be identified.  This is a basis on which to build our approximate locality.  For example, we can make the following obvious approximations:
\begin{align}
\frac{\partial \phi(u,v)}{\partial u}& \approx \frac{\phi(u,v)-\phi(u-a,v)}{a}, \\
\frac{\partial \phi(u,v)}{\partial v}& \approx \frac{\phi(u,v)-\phi(u,v-a)}{a},
\end{align}
from which we can find the d'Alambertian at the point $(u,v)$:
\begin{equation}
\Box \phi(u,v)= \frac{\partial^2\phi(u,v)}{\partial u \partial v} \approx \frac{\phi(u,v)-\phi(u-a,v)-\phi(u,v-a)+\phi(u-a,v-a)}{a^2}.
\end{equation}

From this, the classical dynamics of the scalar field can be approximated on the lattice.  All this stems from the easy identification of the small set of sites $(u-a,v)$, $(u,v-a)$ and $(u-a,v-a)$ as neighbours to $(u,v)$, along with the distances between them, which was quite trivial to do here.  But, of course, we have sacrificed Lorentz invariance in this discretisation scheme.

For a causal set discretisation of the same space (\ie a typical
sprinkling into 2D Minkowski), how might we identify the nearest
neighbours?  Consider an element $x$ sprinkled at co-ordinates
$(u_x,v_x)$.  From the causal set structure alone (no ``cheating''
by directly using the geometry of the Minkowski space we have
sprinkled it into), how do we identify the nearest elements to the
past of $x$?  From the discussion of timelike distance above, we
might select the elements that have the fewest elements causally
between themselves and $x$.  An element $y$ obeying $|I(y,x)|=0$,
where $I(y,x)=\{z \in \cC : y\prec z \prec x \}$ ($\cC$ being the
causal set), clearly fits the bill, and we say that $y$ is
\textit{linked} to the past of $x$ (these links are the relations
not implied by transitivity, that are drawn as lines in a Hasse
diagram).  The problem is that, in a sprinkling of Minkowski,
there are infinitely many such elements with probability 1
\cite{Moore:1988,Bombelli:1988qh}.

\begin{figure}[ht]
\centering
\resizebox{4in}{2in}{\includegraphics{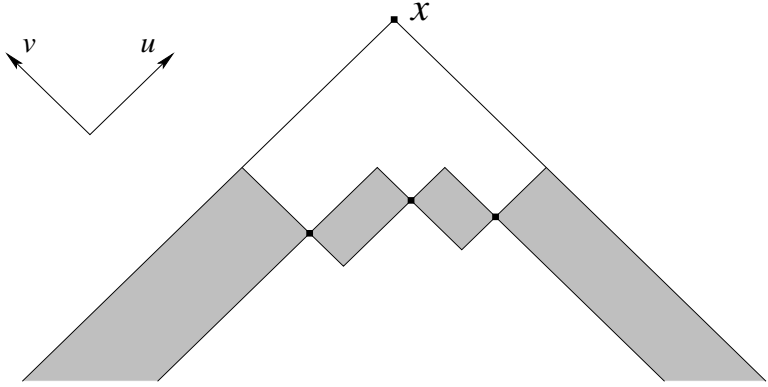}}
\caption{\small{
A spacetime diagram with 4 sprinkled points.  Consider a sprinkling of 2D Minkowski.  Assume that the embedded elements shown to the past of $x$ are linked to $x$ in such a sprinkling.  In order for there to be any more elements linked to $x$, some elements must be sprinkled spacelike to these three, in the shaded region.  But this region extends over infinite volume, and so with probability 1 there will be some elements sprinkled there.
}\label{f:links}}
\end{figure}

To see why in our 2D example, consider a finite set of elements,
all linked to the past of $x$, as shown in figure \ref{f:links}.
From these elements, there is one with a maximal $u$-co-ordinate
$u_{\text{max}}$, and one with a minimal $v$-co-ordinate
$v_{\text{min}}$.  The region for which $u_{\text{max}}<u<u_x$,
$v<v_{\text{min}}$ is spacelike to these elements, and to the past
of an element $x$.  But this region is of infinite extent, so it
is with probability $1$ that at least one element is sprinkled
there, and so there must be at least one more element linked to
$x$.  This process clearly cannot stop at a finite number.  The
elements linked to $x$ are consequently infinite in number and
sprinkled close to the light-cone, back to past infinity.  It can
be seen without much difficulty that this is true also in higher
dimensions.

So we have an infinite set of nearest neighbours (or at least a
very large one in realistic cosmological scenarios), and not much
idea of the geometry between them.  This gives an idea of why
locality is harder to recover for causal sets than for
lattice-like discrete structures.  Once discreteness has been
decided on, it seems that there is a straight choice between the
easy recovery of locality, and Lorentz invariance.  It should
however be commented that Lorentz invariance has been tested to
near the Planck scale, whereas locality has not been tested to
anything near the Planck length.  Thus we might say that Lorentz
invariance is better tested than locality.

Despite these difficulties, it is possible to approximate the d'Alambertian in Minkowski space of even dimension, with reference only to a sprinkled causal set with a scalar field defined on it.  Previous attempts to do this \cite{Daughton:1993,Sorkin:1990bh} rested on the fact that the Green's functions of the d'Alambertian in 2 and 4 dimensions are expressible in terms of causal relations, giving them easy causal set discretisations.  This technique can be used to model the propagation of massless fields from a source to detector, confirming that it is possible for waves to travel on a causal set background without contradicting present observations \cite{Dowker:2005}.  These Green's functions can also be inverted in an attempt to define the d'Alambertian.  However, the resulting d'Alambertian is an unstable approximation, and not much use for evolving a field when the source is not given (although, strangely, this poses no problem when the field is to be calculated from the source).  However, the jury is still out on some revised versions of this approximation method, and they are currently being checked on computer by Rob Salgado.

Recently, another method has been developed which has some analytical backing and has survived the computational tests which previous ideas failed.  This interesting new technique is described here.

\subsection{A non-local, causal approximation of the d'Alambertian}

The first step in finding a causal set approximation to the d'Alambertian is to introduce a \textit{continuum} operator which approximates to the d'Alambertian $\Box \phi(u,v)$, but is non-local, being in the form of a weighted integral over $\phi$ in the past of the point $(u,v)$.  It will then be seen that this operator can be discretised on the causal set.

In a sense, the lattice discretisation given above is also of this type.  In that case the weighted ``integral'' only gives non-zero weight to 4 points: a positive one to $\phi(u,v)$ and $\phi(u-a,v-a)$, and a negative one at $\phi(u-a,v)$ and $\phi(u,v-a)$.  But as has been repeatedly emphasised, we are looking for a Lorentz invariant discretisation.  The following weighted integral can be seen as a ``smearing'' of some such 2D lattice discretisation over all frames.

\begin{equation}
\label{e:box}
\Box \phi(0) \approx k\phi(0) - 2k^2\int_{x \prec 0} d^2x \; \phi(x) \, e^{-kV(x)} \, \Big( 1 -2kV(x) + \frac{1}{2} k^2 V(x)^2 \Big) \\
\end{equation}

For convenience of notation the operator's value at the origin 0
is given.  Here, $V(x)$ is the volume of the ``interval'' region
causally between $x$ and 0, while $k$ parameterises the
non-locality scale ($1/\sqrt{k}$ is taken to be some length above
the Planck scale, but below the known limits on locality).

This continuum approximation to the d'Alambertian is interesting
in its own right, constituting a Lorentz invariant way to
introduce non-locality.  For our purposes, it is important to note
that there is nothing in eqn.(\ref{e:box}) that cannot easily be
discretised on a causal set sprinkled into the Minkowski space. In
the discretised version, the integral becomes a sum over all
elements to the past of the element chosen to be the ``origin'' 0.
The only other thing we need is the volume $V(x)$, which, as
already noted, is approximated by the number of elements causally
between $x$ and 0.  To make a good approximation, the length
associated to the locality scale, $1/\sqrt{k}$, must be
sufficiently far above the length scale set by the fundamental
discreteness, $1/\sqrt{\rho}$.

It may seem surprising that the operator given in
eqn.(\ref{e:box}) does not always give divergent values for
practical situations. As with any approximation, field
configurations can be concocted for which (\ref{e:box}) gives bad
values (for example, a field configuration proportional to the
kernel of the integral in (\ref{e:box})).  The integral in
(\ref{e:box}) is sampling the value of the field over an infinite
volume.  It seems unlikely, on the face of it, that this method
could produce stable answers. How could the value of this operator
avoid being grossly affected by the values of the field ``far down
the light-cone''?  This intuition had put researchers off even
trying to develop a technique like this in the past, but somewhat
miraculously (\ref{e:box}) does indeed give good answers for a
wide variety of practically applicable circumstances, for instance
for plane waves and their superpositions (\ie all classical
solutions of the wave equation).  Although a full justification of
the approximation cannot be included here, another broad class of
situations in which the approximation is successful is given, and
the reason for the success is sketched.

\subsection{How it works}

First, we rewrite eqn.(\ref{e:box}).  The integral

\begin{equation}
J(k)=\int_{x \prec 0} d^2x \; \phi(x) \, e^{-kV(x)} \, \Big( 1 -2kV(x) + \frac{1}{2} k^2 V(x)^2 \Big) \\
\end{equation}
can be written
\begin{equation}
J(k)=\Big( 1\,+\,2k\frac{\partial}{\partial k}\,+\,\frac{1}{2}k^2\frac{\partial^2}{\partial k^2} \Big) I(k)\, ,
\end{equation}
where
\begin{equation}
I(k)=\int_{x \prec 0} d^2x \; \phi(x) \, e^{-kV(x)} \; .
\end{equation}
It is crucial to note that
\begin{align}
\label{e:killk1}
\Big( 1\,+\,2k\frac{\partial}{\partial k}\,+\,\frac{1}{2}k^2\frac{\partial^2}{\partial k^2} \Big) & \; \frac{1}{k}\,=\,0, \\
\label{e:killk2}
\Big( 1\,+\,2k\frac{\partial}{\partial k}\,+\,\frac{1}{2}k^2\frac{\partial^2}{\partial k^2} \Big) & \; \frac{1}{k^2}\,=\,0,
\end{align}
so that terms in $I(k)$ of order $-1$ and $-2$ in $k$ do not affect the value of $J(k)$.

From (\ref{e:box}), we have that

\begin{equation}
\label{e:box2}
\Box \phi(0) \approx k\phi(0) - 2k^2 J(k)
\end{equation}

Here, we will be dealing with fields that have support only in a compact region around the origin.  It is also assumed that a frame can be found in which the field is slowly varying in the $(u,v)$ co-ordinates near the light-cone.  A brief calculation answers the question of why the values of the field far down the light-cone do not greatly affect the value of $\Box \phi(0)$.

Writing $I(k)$ in terms of the $(u,v)$ co-ordinates, we have
\begin{equation}
I(k)=\int_{- \infty}^0  du \int_{-\infty}^0 dv \; e^{-kV(x)} \, \phi(u,v) \; .
\end{equation}

Let us work in units in which $1 \ll k^2$ (\ie one unit of length is much larger than the non-locality length $1/\sqrt{k}$).  To answer our question, let us consider the contribution from the region $u=-\infty \, ... \, -1$, $v=-\infty\,  ...\, 0$, which we will call $I_1(k)$.  Since $|u| \geq 1$, the only significant contribution to the integrand comes from $-\frac{1}{k} < v \leq 0$.  Over this region, $\phi(u,v) \approx \phi(u,0)$.

\begin{align}
I_1(k)&=\int_{-\infty}^{-1} du\, \int_{-\infty}^0 dv\, e^{-kuv} \, \bigr( \phi(u,0) \; + \frac{\partial \phi(u,v)}{\partial v} \Biggl \arrowvert_{v=0}v + ... \bigl)
 \\
&= \int_{-\infty}^{-1} \frac{du}{-ku} \phi(u,0) \; + O(\frac{1}{k^2}) +O(e^{-k}).
\end{align}
Using eqns.(\ref{e:killk1},\ref{e:killk2}), we see that the contribution to $J(k)$ from this region is of order $1/k^3$.  Because of this, the contribution to (\ref{e:box2}) is suppressed by a factor of $1/k$.  This shows that the values of the field in the region $u=-\infty \, ... \, -1$, $v=-\infty\,  ...\, 0$ do not significantly affect the value of this approximation to the d'Alambertian at 0.  Similar calculations in the remainder of the past light-cone of the origin serve to more fully justify (\ref{e:box2}) for fields of this type, and superpositions of them.  Computer simulations have also been carried out for the causal set discretised version, with good results.  These, along with other such results in higher (even) dimensions, will be presented in future work \cite{dowker:2006}.

\subsection{Uses and Consequences}

Now that we have one way to recover locality for fields on a causal set, what can be done with it?  One of the first uses might be in phenomenology.  A classical scalar field dynamics can now be defined.  In this case, the intuition behind the ``swerves'' model described in section \ref{s:swerves} seems to be justified: there is indeed some ``temporal non-locality'', but the influence of the past on the field at a point is limited to within a certain proper time from that point.  It would be interesting to set up computer simulations to see if wave packets in this field do indeed swerve away from geodesic trajectories.  Perhaps other non-standard effects could also be identified in this new field dynamics.  It would also be an interesting exercise to find a way to quantise the field, and look for similar results there.

One of the most intriguing consequences is for causal set dynamics, as mentioned in section \ref{s:actions}.  The approximation to the d'Alambertian is conjectured to hold good for curved spacetimes also, and hopefully it will be possible to extend arguments like those of the previous section to this case.  But how can this help us to find an action for causal sets?  Consider the field $\Box \sigma(0,x)$, where $\sigma(x,y)$ is Synge's world function (\ie half of the square of the geodesic distance between $x$ and $y$).  It can be seen from some of the results in \cite{Synge:1960} that the d'Alambertian of \textit{this} field at the origin gives the scalar curvature there:

\begin{equation}
R(0)=\Box \, \Box \, \sigma(0,x)\Biggl \arrowvert_{x=0}.
\end{equation}

The geodesic length between two timelike points in a causal set can be estimated (independently, it is conjectured, of curvature).  Therefore, if we have a way of estimating the d'Alambertian of fields in curved space times, we also have a way of estimating the scalar curvature.  If this method turns out to be correct, and the values found are stable and practically calculable, it will be of great significance for causal set dynamics.

This result is, hopefully, only the first handle on the problem of
locality in causal sets, and consideration of what has been learnt
may lead to the development of more techniques, as the reason for
this success is more fully grasped.  For instance, in one of the
original articles on what became the causal set program, Myrheim
suggests a way to find components of the Ricci tensor from causal
information alone \cite{Myrheim:1978ce}.  Just as the above result
can be seen as a smearing of a lattice d'Alambertian over the
whole light-cone, it may be possible that Myrheim's
direction-dependent results could be smeared over the light-cone
to get another way of estimating the scalar curvature.  The
ultimate goal would be would be to find an expression for the
action which is combinatorially simple and compelling, and which
gives sensible values for non-manifoldlike causal sets.  Work on
this topic has only just begun.

\section{Conclusions}

Discreteness provides a solution for many of the problems we confront in our attempts to construct a theory of quantum gravity.  From the assumptions of discreteness and standard Lorentz invariance, we find that our choices of fundamental histories are extremely limited.  Although this should not discourage other attempts to reconcile the two
\footnote{
See for example \cite{Kempf:2003qu}.
}
, it has been argued here that, at present, the causal set is the only proposal that does so.

The causal set program is an active and growing one.  Many projects in progress have not been mentioned above.  Attempts to identify the ``atoms'' that carry the Black hole entropy have been made \cite{Dou:2003af,Dou:1999fw}, and this work is currently being extended to higher dimensions by Fay Dowker and Sara Marr.  Further work on the question of ``observables'' has also been carried out \cite{Major:2005fy}.  Pros and cons of an amplitude-based dynamics are also being investigated.  As well as this work, the basic causal set idea continues to inspire other approaches \cite{Raptis:2005jw,Isham:2003dk}.

Every statement of a result given here raises many more questions,
only some of which are being pursued.  This multiplicity of
unanswered questions, the relatively small set of prerequisites
needed to contribute to them, and the comparative, ``strategic''
perspective on quantum gravity that the approach offers, make
causal sets an attractive field for both new and experienced
researchers.

\section*{Acknowledgements}

The author would like to thank to David Rideout and Rafael Sorkin
for helpful comments.

\bibliographystyle{h-physrev3}
\bibliography{refs}
\end{document}